\documentclass{appolb}
\usepackage{epsfig}

\begin{document}
\title{CP VIOLATION IN CHARGINO PRODUCTION\\ IN $e^+e^-$
COLLISIONS\thanks{Presented by K.~Rolbiecki at the XXXI
International Conference of Theoretical Physics, ``Matter to the
Deepest", Ustro\'{n}, Poland,
September 5--11, 2007.}%
}
\author{Krzysztof Rolbiecki \and Jan Kalinowski\thanks{The authors are supported by the
Polish Ministry of Science and Higher Education Grant
No.~1~P03B~108~30 and by the EU  Network MRTN-CT-2006-035505 ``Tools
and Precision Calculations for Physics Discoveries at Colliders".
 }
\address{Institute of Theoretical Physics, University of Warsaw\\
Ho\.{z}a 69, 00-681 Warsaw, Poland}} \maketitle
\begin{abstract}
We present the analysis of CP-violating effects in non-diagonal
chargino pair production in $e^+e^-$ collisions. These effects
appear only at the one-loop level. We show that CP-odd asymmetries
in chargino production are sensitive to the phases of $\mu$ and
$A_t$ parameters and can be of the order of a few \%.
\end{abstract}

\section{Introduction}
Supersymmetry (SUSY) is one of the most promising extensions of the
Standard Model (SM) \cite{MSSM} since, among other things, it solves
the hierarchy problem, provides a natural candidate for dark matter
\textit{etc}. It also introduces many new sources of CP violation
that may be needed to explain baryon asymmetry of the universe.
These phases, if large $\mathcal{O}(1)$, can cause problems with
satisfying experimental bounds on lepton, neutron and mercury
EDMs~\cite{susycp}. This can be overcome by pushing sfermion spectra
above a TeV scale or arranging internal cancelations
\cite{Ibrahim:2007fb}.

Most unambiguous way to detect the presence of CP-violating phases
would be to study CP-odd observables measurable at future
accelerators --- the LHC and the ILC. Such observables in the
chargino sector are, for example, triple products of momenta of
initial electrons, charginos and their decay products
\cite{Kittel:2004kd}. However they require polarized initial
electron/positron beams or measurement of chargino polarization.

In this talk we present another possibility of detecting
CP-violating phases in the chargino sector. As it was recently
pointed out \cite{Osland:2007xw,my}, in non-diagonal chargino pair
production
\begin{equation}
e^+e^-\to\tilde{\chi}_1^\pm\tilde{\chi}_2^\mp \label{produkcja}
\end{equation}
a CP-odd observable can be constructed beyond tree-level from
production cross section without polarized $e^+e^-$ beams or
measurement of chargino polarization. We show here the results of
the full one-loop calculation of this effect. In the
reaction~(\ref{produkcja}) the CP violation can be induced by the
complex higgsino mass parameter $\mu$ or complex trilinear coupling
in top squark sector $A_t$. Since these asymmetries can reach a few
percent, they can be detected in simple event-counting experiments
at future colliders.

\section{CP-odd asymmetry at one loop}

In $e^+e^-$ collisions charginos are produced at tree-level via the
$s$-channel $\gamma,Z$ exchange and $t$-channel $\tilde{\nu}_e$
exchange. As it was shown in \cite{Choi:1998ei} no CP violation
effects can be observed at the tree-level for the production
processes of diagonal $\tilde{\chi}_i^+ \tilde{\chi}_i^-$ and
non-diagonal $\tilde{\chi}_i^+ \tilde{\chi}_j^-$ chargino pairs
without the measurement of polarization of final chargino. However
the situation is different for non-diagonal production if we go
beyond tree-level approximation.

Radiative corrections to the chargino pair production include the
following generic one-loop Feynman diagrams: the virtual vertex
corrections, the self-energy corrections to the $\tilde{\nu}$, $Z$
and $\gamma$ propagators, and the box diagrams contributions. We
also have to include corrections on external chargino legs.

One-loop corrected matrix element squared is given by
\begin{eqnarray}
|\mathcal{M}_{\mathrm{loop}}|^2 = |\mathcal{M}_{\mathrm{tree}}|^2 +
2 \mathrm{Re}(\mathcal{M}_{\mathrm{tree}}^*
\mathcal{M}_{\mathrm{loop}} )\, .
\end{eqnarray}
Accordingly, the one-loop CP asymmetry for the non-diagonal chargino
pair is defined as
\begin{eqnarray}
&& A_{12}=\frac{\sigma^{12}_{\rm loop}-
  \sigma^{21}_{\rm loop}}{\sigma^{12}_{\rm tree}+
  \sigma^{21}_{\rm tree}}\, ,
  \label{CPasym}
\end{eqnarray}
where $\sigma^{12}$, $\sigma^{21}$ denote cross sections for
production of $\tilde{\chi}_1^+ \tilde{\chi}_2^-$ and
$\tilde{\chi}_2^+ \tilde{\chi}_1^-$, respectively. Since the
asymmetry vanishes at tree-level it has to be finite at one loop,
hence no renormalization is needed.

\begin{figure}[!t]
\begin{center}
\includegraphics[scale=0.9]{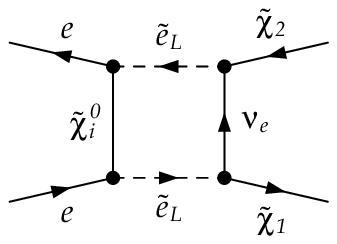}\hskip 1cm
\includegraphics[scale=0.35]{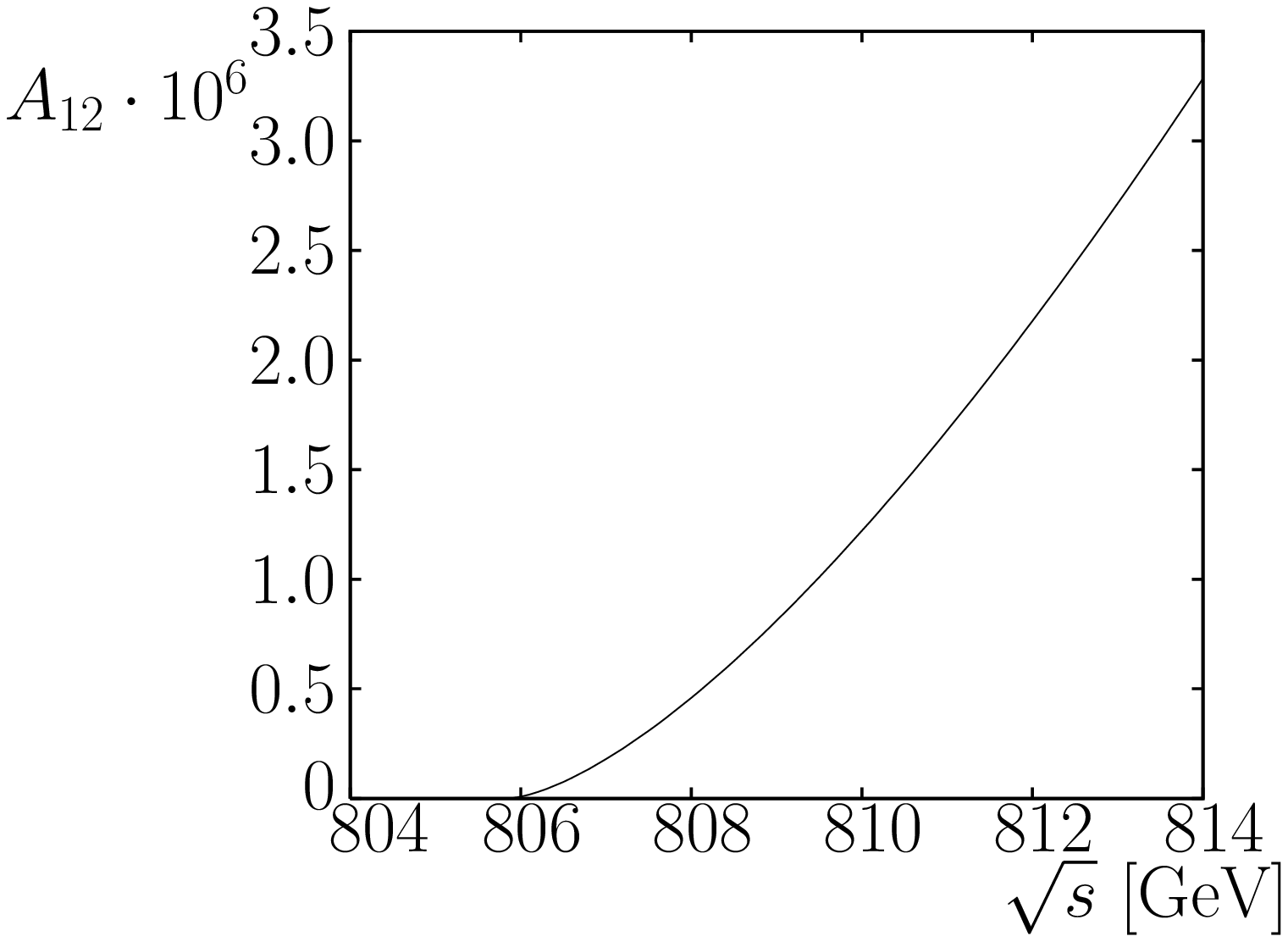}
\end{center}
\caption{Box diagram with selectron exchange and its contribution to
the asymmetry $A_{12}$ vs.\ center of mass energy. The selectron
mass is 403~GeV.\label{fig:thrs_scan}}
\end{figure}

The CP asymmetry Eq.~(\ref{CPasym}) arises due to the interference
between complex couplings, which in our case are due to complex
mixing matrices of charginos or stops, and non-trivial imaginary
part from Feynman diagrams --- the absorptive part. Such
contributions appear when some of the intermediate state particles
in loop diagrams go on-shell. This is illustrated in
Fig.~\ref{fig:thrs_scan} where the contribution to $A_{12}$ from
double selectron exchange appears at the threshold for selectron
pair production at $\sqrt{s}=806$~GeV.

\section{Numerical results}

For the numerical results in this section we use two parameter sets
(A) and (B) with gaugino/higgsino mass parameters defined as follows
at the low scale:
\begin{eqnarray}
&\mbox{A:}&\quad |M_1| = 100\mbox{ GeV},\quad M_2 = 200\mbox{ GeV},
\quad|\mu| = 400\mbox{ GeV},\nonumber \\
&\mbox{B:}&\quad |M_1| = 250\mbox{ GeV}, \quad M_2 = 200\mbox{ GeV},
\quad |\mu| = 300\mbox{ GeV},\nonumber
\end{eqnarray}
and with $\tan\beta=10$. This gives the following chargino masses:
\begin{eqnarray*}
&\mbox{A:}&\quad m_{\tilde{\chi}^-_1} = 186.7 \mbox{ GeV},\quad
m_{\tilde{\chi}^-_2} = 421.8 \mbox{ GeV},\\
&\mbox{B:}&\quad m_{\tilde{\chi}^-_1} = 175.6 \mbox{ GeV},\quad
m_{\tilde{\chi}^-_2} = 334.5 \mbox{ GeV}.
\end{eqnarray*}
For the sfermion mass parameters in scenario (A) we assume
\begin{eqnarray*}
&&m_{\tilde{q}}\equiv M_{\tilde{Q}_{1,2}}=M_{\tilde{U}_{1,2}}=M_{\tilde{D}_{1,2}}=450\mbox{ GeV},\nonumber\\
&&M_{\tilde{Q}}\equiv M_{\tilde{Q}_{3}}=M_{\tilde{U}_{3}}=M_{\tilde{D}_{3}}=300\mbox{ GeV},\\
&&m_{\tilde{l}}\equiv
M_{\tilde{L}_{1,2,3}}=M_{\tilde{E}_{1,2,3}}=150\mbox{ GeV},
\end{eqnarray*}
and for the sfermion trilinear coupling:
$|A_{t}|=-A_{b}=-A_{\tau}=A=400\mbox{ GeV}$. Scenario (B) is for
comparison with Ref.~\cite{Osland:2007xw} for which we take
$$M_{S}= M_{\tilde{Q}}= M_{\tilde{U}} =M_{\tilde{D}} =M_{\tilde{L}}=
M_{\tilde{E}}=10\mbox{ TeV}.$$

\begin{figure}[!t]
\begin{center}
\includegraphics[scale=0.365]{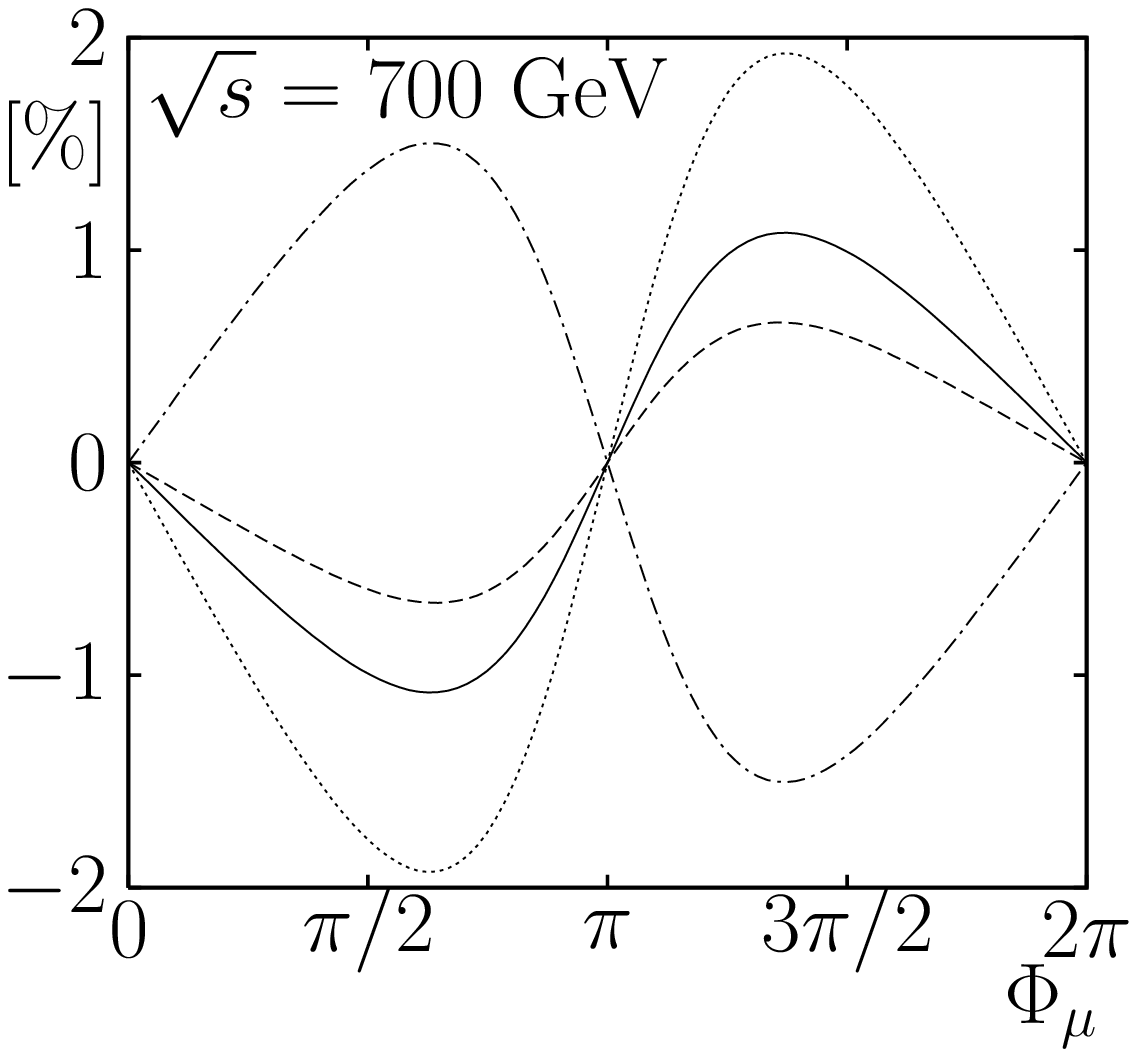}
\includegraphics[scale=0.35]{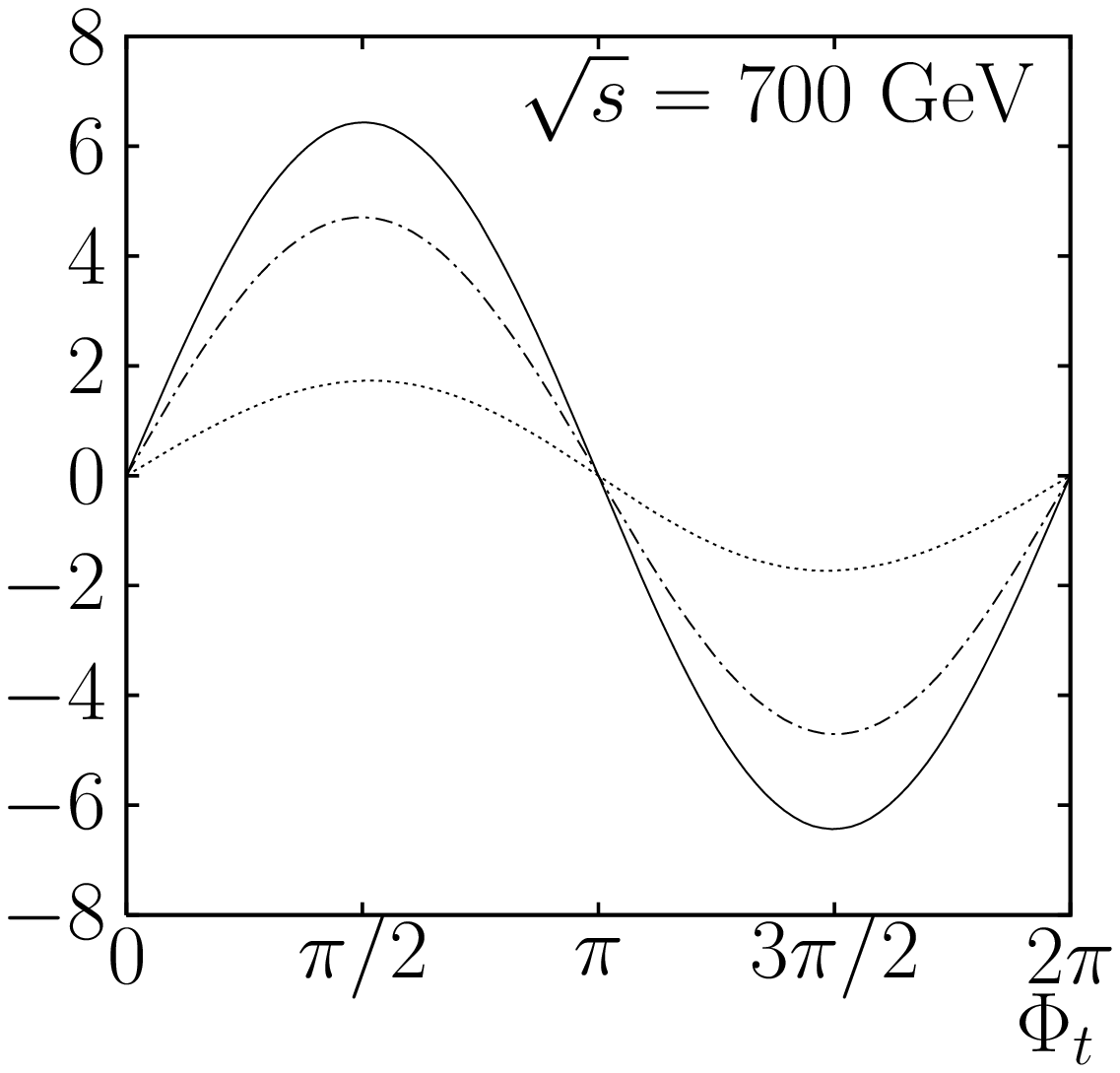}
\includegraphics[scale=0.365]{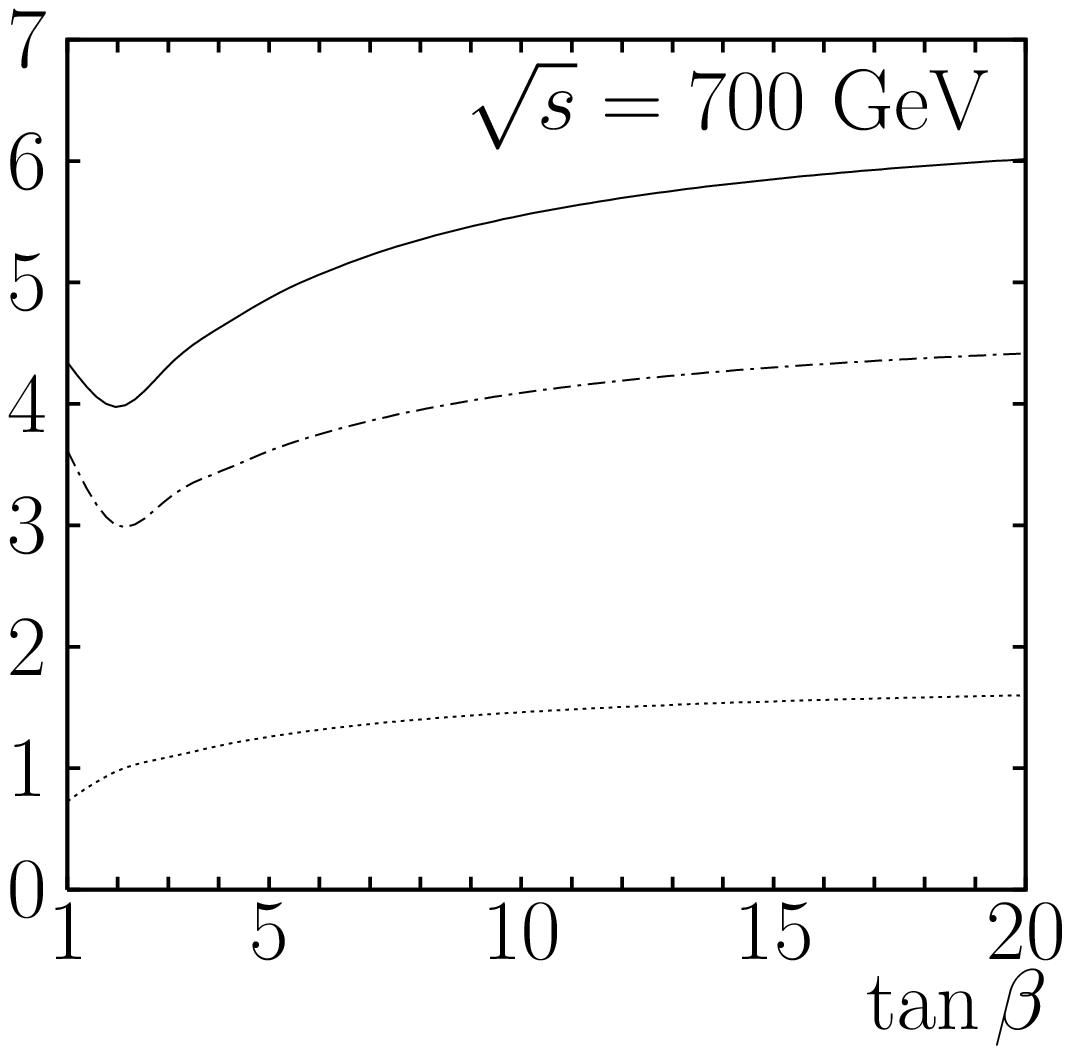}
\end{center}\vspace{-0.1cm}
\caption{Asymmetry $A_{12}$ in scenario (A) as a function of the
phase of $\mu$ parameter (left), the phase of $A_t$ (middle), and as
a function of $\tan\beta$ with $\Phi_t=\pi/3$ (right). Different
lines denote full asymmetry (full line) and contributions from box
(dashed), vertex (dotted) and self energy (dash-dotted) diagrams.
\label{fig2}}
\end{figure}

In our numerical analysis we consider the dependence of the
asymmetry (\ref{CPasym}) on the phase of the higgsino mass parameter
$\mu = |\mu| e^{i \Phi_\mu}$ and soft trilinear top squark coupling
$A_t = |A_t| e^{i \Phi_t}$. In Fig.~\ref{fig2} we show the CP
asymmetry in scenario (A) as a function of the phase of $\mu$ and
$A_t$, left and middle panel, respectively.  Contributions due to
box corrections, vertex corrections and self energy corrections have
been plotted in addition to the full result. In this scenario the
asymmetry can reach $\sim 1\%$ for the $\mu$ parameter and $\sim
6\%$ for $A_t$, respectively. We note that for the asymmetry due to
the non-zero phase of the higgsino mass parameter there are
significant cancelations among various contributions. In addition,
we also show in the right panel of Fig.~\ref{fig2} the dependence of
the asymmetry due to $A_t$ as a function of $\tan\beta$.

For the asymmetry generated by the $\mu$ parameter all possible
one-loop diagrams containing absorptive part contribute. The
situation is different for the phase of  the trilinear coupling
$A_t$ --- when chargino mixing matrices remain real. In this case
only vertex and self-energy diagrams containing stop lines
contribute to the asymmetry~\cite{my}.

\begin{figure}[!t]
\begin{center}
\includegraphics[scale=0.335]{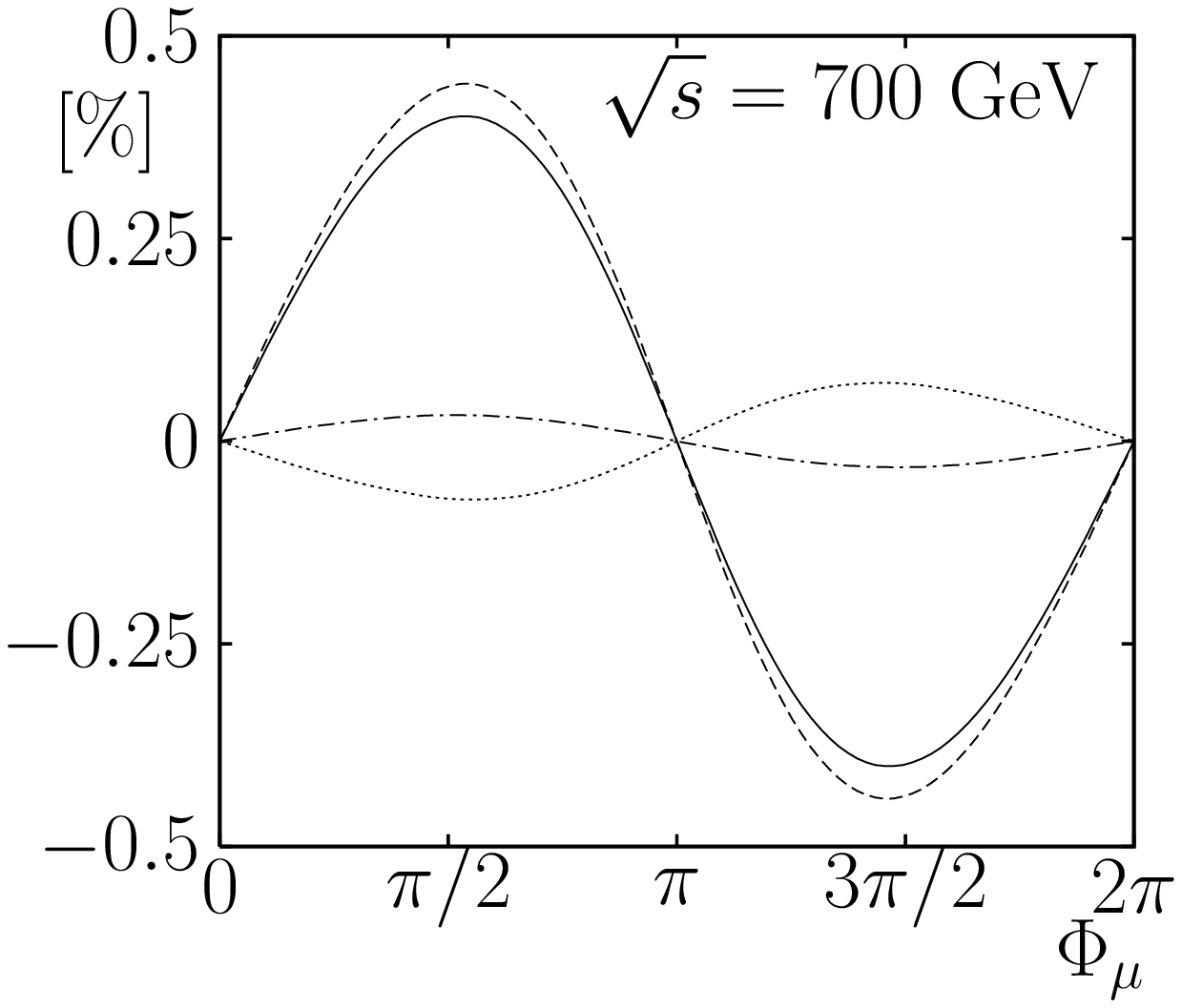}
\includegraphics[scale=0.32]{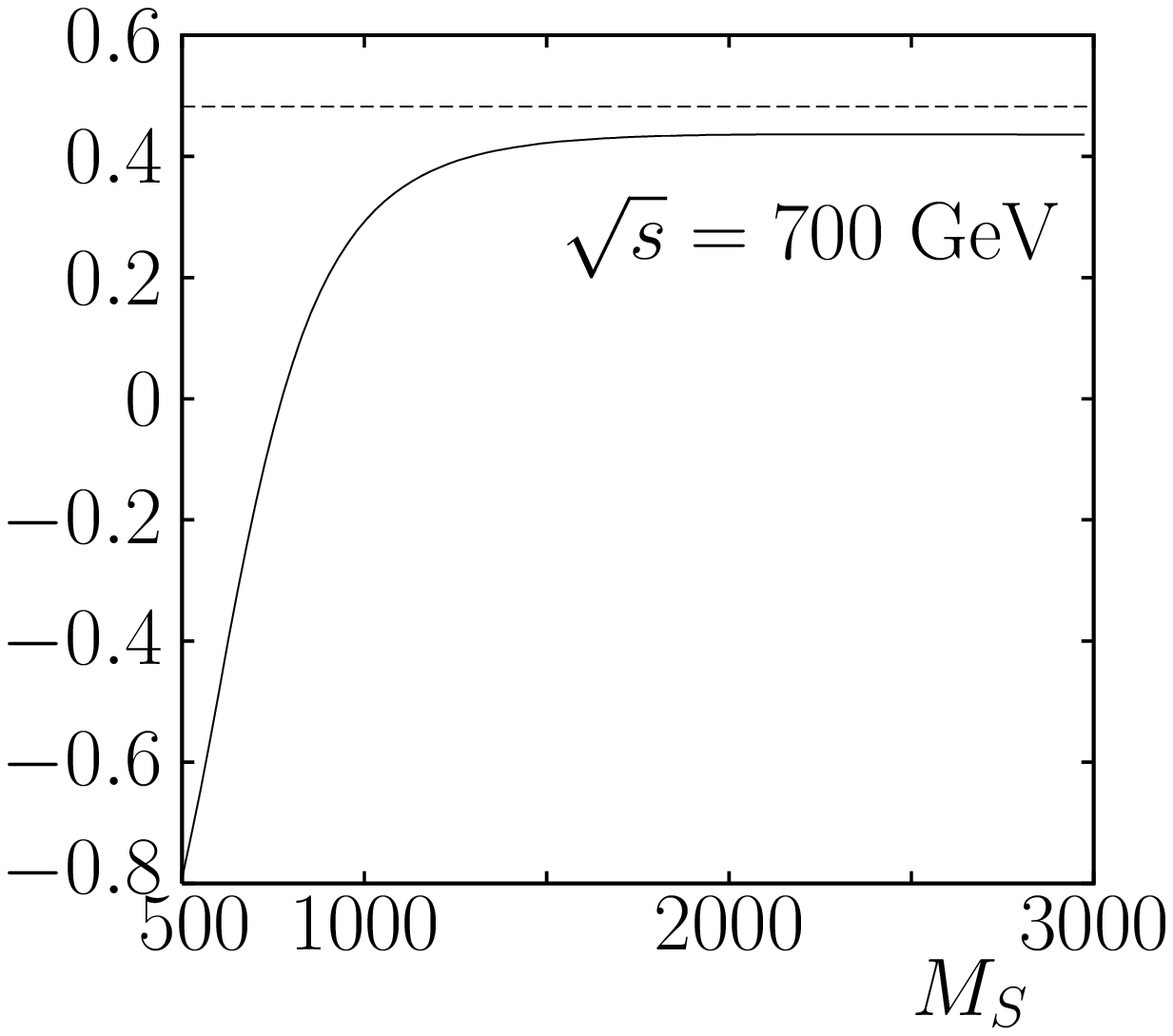}
\includegraphics[scale=0.32]{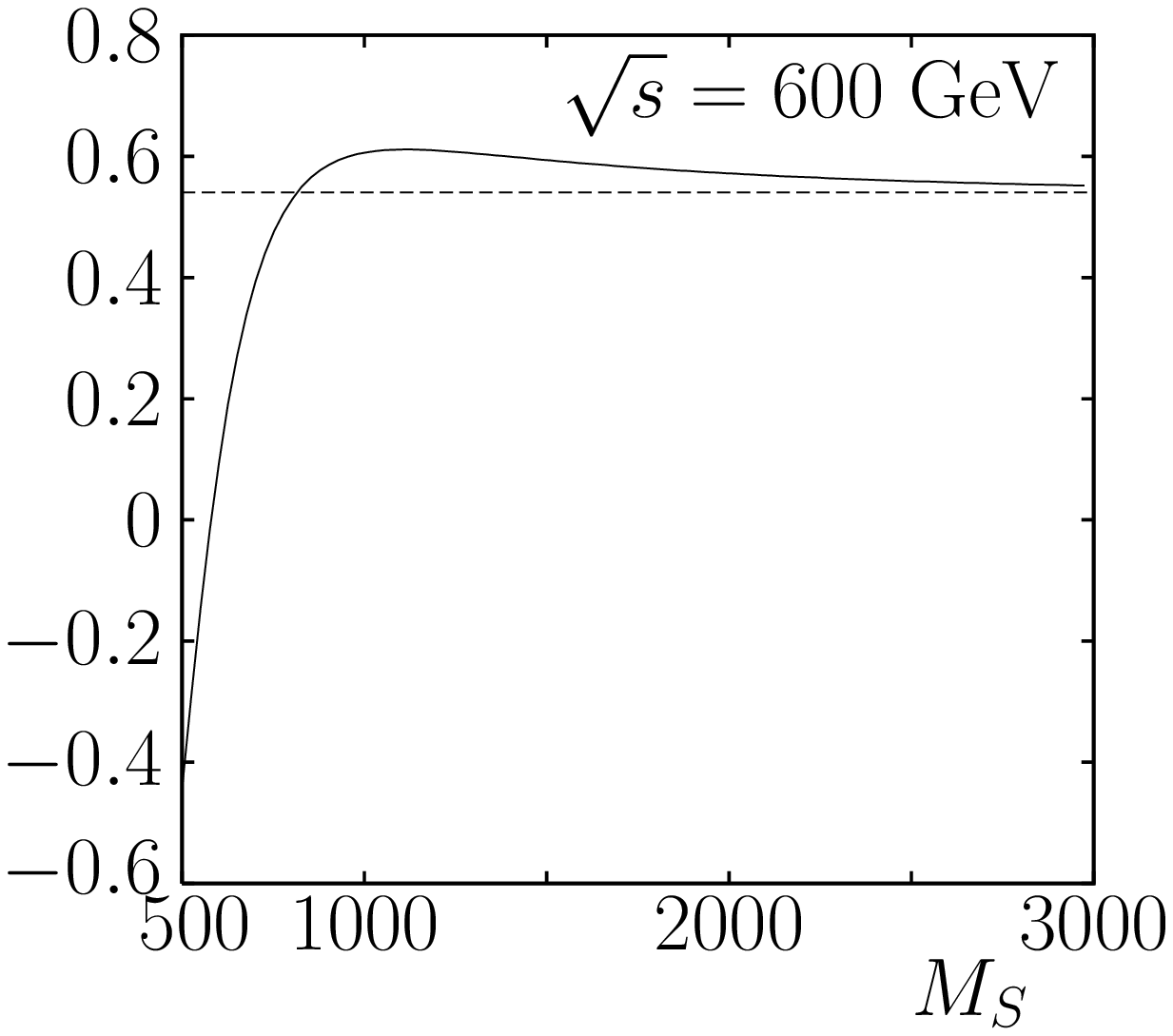}
\end{center}
\caption{Left: Asymmetry $A_{12}$ in scenario (B) as a function of
the phase $\Phi_\mu$. Different lines denote full asymmetry (full
line) and contributions from box (dashed), vertex (dotted) and self
energy (dash-dotted) diagrams. Middle and Right: Asymmetry $A_{12}$
as a function of the universal scalar mass $M_S$ with
$\Phi_\mu=\pi/2$ at different cms. The full lines denote full result
and dashed lines show only the box contributions after neglecting
diagrams with slepton exchange.\label{fig3}}
\end{figure}

We present also the results for the heavy sfermion scenario~(B).
This is to compare with~\cite{Osland:2007xw} where only box diagrams
with $\gamma$, $W$, $Z$ exchanges have been calculated neglecting
all sfermion contributions. As can be seen in the left panel of
Fig.~\ref{fig3} these gauge-box diagrams constitute the main part of
the asymmetry $A_{12}$, however this is due to partial cancelation
of vertex and self-energy contributions. For lower values of the
universal scalar mass $M_S$ the discrepancy between full and
approximate result of~\cite{Osland:2007xw} increases significantly.
This is illustrated in the middle and right panel of Fig.~\ref{fig3}
where we show two paths of approaching of the full result to the
gauge-box approximation as the function of $M_S$. As can be seen
these paths depend strongly on the center of mass energy.

\section{Summary}
It has been shown that CP-odd asymmetry can be generated in
non-diagonal chargino pair production with unpolarized
electron/positron beams. The asymmetry is pure one-loop effect and
is generated by interference between complex couplings and
absorptive parts of one loop integrals. The effect is significant
for the phases of the higgsino mass parameter $\mu$ and the
trilinear coupling in stop sector $A_t$. At future linear collider
it may give information about CP violation in chargino and stop
sectors.


\begin{thebibliography}{100}

\bibitem{MSSM}
  H.~P.~Nilles,
  Phys.\ Rept.\ {\bf 110} (1984) 1;
  H.~E.~Haber and G.~L.~Kane,
  Phys.\ Rept.\  {\bf 117} (1985) 75.

\bibitem{susycp}
  V.~Barger, T.~Falk, T.~Han, J.~Jiang, T.~Li and T.~Plehn,
  Phys.\ Rev.\ D {\bf 64} (2001) 056007
  [arXiv:hep-ph/0101106].

\bibitem{Ibrahim:2007fb}
  T.~Ibrahim and P.~Nath,
  arXiv:0705.2008 [hep-ph] and references therein.

\bibitem{Kittel:2004kd}
  O.~Kittel, A.~Bartl, H.~Fraas and W.~Majerotto,
  Phys.\ Rev.\  D {\bf 70} (2004) 115005
  [arXiv:hep-ph/0410054].

\bibitem{Osland:2007xw}
  P.~Osland and A.~Vereshagin,
  Phys.\ Rev.\  D {\bf 76} (2007) 036001
  [arXiv:0704.2165 [hep-ph]].

\bibitem{my}
  K.~Rolbiecki and J.~Kalinowski,
  arXiv:0709.2994 [hep-ph].

\bibitem{Choi:1998ei}
  S.~Y.~Choi, A.~Djouadi, H.~S.~Song and P.~M.~Zerwas,
  Eur.\ Phys.\ J.\  C {\bf 8} (1999) 669
  [arXiv:hep-ph/9812236].



\end{thebibliography}
\end{document}